\documentclass[%
 aip,
 amsmath,amssymb,
 reprint,%
rsi
]{revtex4-2}

\usepackage{graphicx}
\usepackage{bm}

\usepackage[utf8]{inputenc}
\usepackage[T1]{fontenc}

\usepackage[colorlinks=true,linkcolor=blue,citecolor=blue,urlcolor=blue,pdfborder={0 0 0}]{hyperref}
\usepackage{siunitx}
\usepackage{physics}

\usepackage[normalem]{ulem} 

\newcommand{\rs}{\scriptscriptstyle} 
\newcommand{\todo}[1]{{\color{red}#1}}

\newcommand{\VIset}{{V_{I_\mathrm{\rs set}}}}
\newcommand{\VIsens}{{V_{I_\mathrm{\rs sens}}}}
\newcommand{\LV}{{\mathrm{LV}}}
\newcommand{\LVp}{{\mathrm{LV}_+}}
\newcommand{\LVm}{{\mathrm{LV}_-}}
\newcommand{\HV}{{\mathrm{HV}}}
\newcommand{\HVp}{{\mathrm{HV}_+}}
\newcommand{\HVm}{{\mathrm{HV}_-}}
\newcommand{\Cp}{{\mathrm{C}_+}}
\newcommand{\Cm}{{\mathrm{C}_-}}
\newcommand{\Iscale}{\SI{4.3}{\volt/\ampere}}
\newcommand{\VGSth}{{V_\mathrm{GS(th)}}}

\begin{document}


\title{Fast Magnetic Coil Controller for Cold Atom Experiments} 



\author{L. Uhthoff-Rodríguez}
\author{A. Hernández-López}
\author{E. G. Alonso-Torres}
\author{E. Esquivel-Ramírez}
\author{G. Carmona-Torres}
\author{\\C. Gardea-Flores}
\author{J. A. Seman}
\author{A. Paris-Mandoki}
\email{asaf@fisica.unam.mx}
\affiliation{Instituto de Física, Universidad Nacional Autónoma de México, Ciudad de México C.P. 04510, Mexico}

\date{\today}

\begin{abstract}



Cold atoms experiments employ magnetic fields, commonly generated by coils, as an essential tool to control and manipulate atomic samples. In these experiments, it is often necessary to rapidly switch the magnetic field between two values. However, typical power supplies have a limited switching time for the current flowing through the coil. We present a control scheme implemented as an electronic circuit that overcomes this limitation. A faster control is achieved by momentarily applying an on-demand high voltage when the control signal variation surpasses the limits that the conventional power supply can follow, allowing a faster current flow into the magnetic coil. In our specific application, a coil with inductance of \SI{491}{\micro\henry} and resistance of \SI{0.26}{\ohm}, corresponding to a time constant of $\sim\SI{1.9}{\milli\second}$, is driven so that its current faithfully follows the control signals from $\SI{-1}{\ampere}$ to $\SI{+1}{\ampere}$. Capable of completing a full-scale transition in just $\sim\SI{31}{\micro\second}$, with an effective bandwidth of $\SI{15.2}{\kilo\hertz}$. This corresponds to an improvement by a factor of more than 20 in both bandwidth and switching speed over a conventional power supply. By appropriately selecting the components of the circuit, both the bandwidth and the switching time can be tuned to match specific needs within a wide range of inductive and power requirements.
\end{abstract}


\maketitle 

\section{Introduction}
Cold atom experiments stand out for the breathtaking level of control they allow. Indeed, these setups have made it possible to manipulate quantum degenerate bosonic and fermionic systems~\cite{blochUltracoldGases2008} as well as  individual atoms~\cite{schlosserLoadingSingleAtoms2001, kaufmanTwizzerArraysUltracoldAtoms2021, shersonSingleAtomMottInsulator2010}, ions \cite{neuhauserFirstSingleIon1980, winelandSingleMgIon1981, zipkesSingleIonBoseCondensate2010}, and photons \cite{changQuantumNonlinearOptics2014}. To achieve this sophisticated degree of control, one of the most important tools is magnetic fields, since they play a central role in many of the techniques employed for creating and manipulating the cold atomic samples~\cite{Metcalf1999,bergemanMagnetostaticTrappingFieldsNeutralAtoms, phillipsFirstObservationTrappedNeutralAtoms, esslingerMagneticTransportTrappedColdAtoms, happerOpticalPumping1972, chengFeshbackResonancesUltracolGases2010,Grier2013a}.

In particular, many of the processes employed in cold atoms experiments require abruptly switchinging the field between two different values. For instance, laser cooling often involves a stage in which atoms are trapped in a magnetic field and then further cooled down after abruptly switching the field off with processes known as optical molasses~\cite{lettOpticalMolasses1989,Metcalf1999, dedmanFastSwitchingMagneticFieldsMOT2001}. Control of interatomic interactions in the ultracold regime by means of a magnetic Fano-Feshbach resonance sometimes requires a very fast sweep (below \SI{100}{\micro\second}) between two values of the magnetic field~\cite{matyjakiewiczProbingFermionicCondensates2008, kellCompactFastCoilFeshback2021}.

Magnetic coils are by far the most widely used method in these experiments, as they offer the possibility of manipulating the intensity and direction of the magnetic field, and even extinguishing it completely. It is often the case in the cold atoms community that home-built circuits are developed to control the current that passes through the coils in a very precise and versatile fashion.~\cite{dedmanFastSwitchingMagneticFieldsMOT2001, ClaussenDynamicsBoseCondensate2003, stummer2007, stummer2011, YangBipolarCurrentDriver2019, MerkelMagFieldStabili2019, kellCompactFastCoilFeshback2021, sabulskyCurrentControllersOptimizing2024}

In practice, coils in cold atom experiments typically have resistances of, at most, a few ohms and inductances of $\SI{0.1}{\milli\henry}-\SI{5}{\milli\henry}$ resulting in $L/R$ time constants of $\SI{0.1}{\milli\second}\sim\SI{5}{\milli\second}$, depending on the application. However, even for the larger coils, switching times of $\sim\SI{0.1}{\milli\second}$ are still desirable to avoid decrease in density or atom loss. The power supplies usually used to drive them are designed for large steady-state currents, have a voltage limit of a few volts, and are optimized for stability rather than speed. As the maximum rate of change of current through a coil is limited by the maximum applied voltage, $V_\mathrm{max}$, and the coil’s inductance $L$, following 
\begin{equation}
\dv{I}{t} = \frac{V_\mathrm{max}}{L}, 
\label{eq:CurrentSpeed}
\end{equation}
fast magnetic field switching requires either a high voltage or a low inductance.

Two strategies can be used to achieve fast switching: i) Reduce the relevant inductance to a few microhenries using an auxiliary coil to add a small offset to a large field from a higher inductance coil~\cite{ClaussenDynamicsBoseCondensate2003, eigenExploringBoseGasesAuxCoils2019} or ii) Increase the applied voltage as much as possible. Fast current turn-off is commonly achieved using snubber circuits, which clamp the high voltage spike generated by the coil to dissipate the energy stored in the magnetic field~\cite{dedmanFastSwitchingMagneticFieldsMOT2001}. Conversely, fast turn-on has been implemented using a large storage capacitor to provide a high-voltage pulse to the magnetic coils~\cite{stummer2007}, a technique also employed in accelerator physics~\cite{AhmedFastMagnet2019}. Moreover, a unipolar, capacitor-based coil driver capable of switching between different current levels was developed to manipulate Feshbach resonances~\cite{ClaussenDynamicsBoseCondensate2003}. For bipolar current control, solutions based on high-power operational amplifiers have been used~\cite{sabulskyCurrentControllersOptimizing2024,YangBipolarCurrentDriver2019}. These are easy to implement but their monolithic nature restricts their customizability. Industrial instruments designed for driving inductive loads, such as the Kepco BOP-L series, are also available. While they are general-purpose devices, better performance can often be achieved with tailor-made solutions.

In this paper we present an electronic circuit that effectively overcomes the difficulty of fast current bidirectional control in magnetic coils with retrofit capability. The presented components have been chosen for the application of quickly changing the current in shim coils used to compensate for external magnetic fields or to set a small bias field. Nonetheless, the circuit can be adapted to other applications by sizing the components accordingly. 

To achieve fast current regulation, a high-voltage and low-current supply is used to charge capacitors, which can then provide a high current at high voltage for short periods of time and complement a steady-state low-voltage and high-current power supply during sudden changes of the current setpoint, shortening the time of the transient behavior. A drawback of this circuit is that it requires a charging period to prepare for fast dynamic changes in the set current. However, this limitation is compatible with most cold-atom experiments, which also require a charging period to collect cooled atoms, where the coil currents are kept constant before executing an experimentation and measurement sequence.

\section{Principle of Operation}

\begin{figure}
    \centering
    \includegraphics[width=\linewidth]{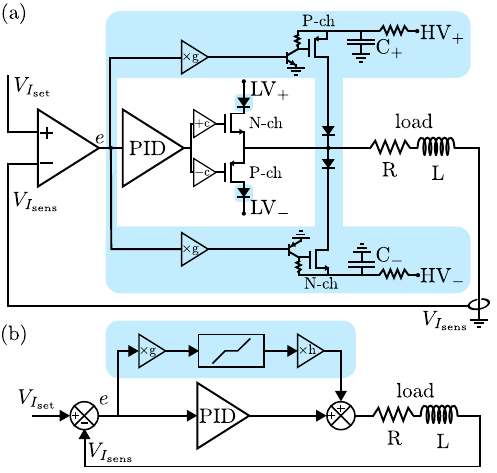}
    \caption{\textbf{Simplified diagrams of current regulator.} The unshaded region shows the basic current regulating circuit and the shaded region includes the additions that enable fast switching. (a) Circuit schematic, (b) block diagram. The triangular symbols with $\times g$ or $\times h$ apply a gain with factor $g$ or $h$ respectively and the triangular symbols with $\pm c$ apply an offset of $c$ to the signal to cancel the gate threshold voltage $\VGSth$ of each MOSFET.}
    \label{fig:circuit}
\end{figure}

A simplified schematic of the circuit is presented in Fig.~\ref{fig:circuit}a. This circuit is powered by two separate power supplies: a ``low voltage'' ($\LV$) one that can output a large current, which drives the load during steady state, and a ``high-voltage'' ($\HV$) supply that is used to charge the capacitors $\Cp$ and $\Cm$ that drive the fast switching part of the circuit. We will begin by describing the basic current regulating circuit (unshaded) and then detail the fast switching section (shaded). The actual value of the current passing through the load is measured with a current sensor that outputs a voltage $\VIsens$ which is compared to the desired set voltage $\VIset$. The resulting error signal $e=\VIset-\VIsens$ is fed into a PID (proportional-integral-derivative) controller that drives a pair of complementary N- and P-channel MOSFETs arranged in a push-pull configuration. In this configuration, when $\VIset > \VIsens$ the output voltage of the PID controller increases causing the conductivity of the N-channel MOSFET to also increase so that more current flows from $\LVp$ to the load until $\VIset \approx \VIsens$. Conversely, for $\VIset < \VIsens$ the PID controller also adjusts its output until $\VIset \approx \VIsens$ by increasing, now, the conductivity of the P-channel MOSFET. Before being fed to the gates of the $\LV$ MOSFETs, the output of the PID is offset by the conduction threshold voltage of each MOSFET ($\VGSth$) so that there is no dead-zone, meaning that they begin conduction as soon as the PID output starts deviating from zero. This offset can even exceed by a small margin the threshold voltage resulting in some power dissipated by a constant bias current at the $\LV$ MOSFETs but at the same time providing some immunity from fluctuations in the threshold voltages. Because the MOSFETs are complementary, only one of them conducts current at any given time, except for the small bias current. As the relationship between the gate voltage and the conductivity of the MOSFETs is not always linear, and it may even be different for the N- and the P-channel MOSFETs,  we employ the PID controller to ensure that the linearity of the circuit is only limited by the linearity of the current sensor. 

The fast switching part of the circuit shown in the shaded part of Fig.~\ref{fig:circuit}a contains two capacitors that are charged to high voltage which can be negative ($\HVm$) or positive ($\HVp$). The discharge of these capacitors towards the magnetic coil load is controlled by an additional pair of MOSFETs, arranged in an effective push-pull configuration. The diodes shown in schematic are there so that the current from the $\HV$ supply does not flow into the $\LV$ supply and vice versa. As opposed to the $\LV$ part of the circuit, the power MOSFETs for the $\HV$ part are driven directly by the error signal and there is no offset added to it so they only become active when the magnitude of the error signal is larger than the MOSFETs' conduction threshold voltage $\VGSth$. As a result, the $\HV$ part of the circuit is inactive when the circuit is in a steady state and the error signal is close to zero. For an abrupt change in the current setpoint $\VIset$, the error signal changes accordingly and allows current to flow from an $\HV$ capacitor to the load. As the voltage of the capacitors is much larger that the $\LV$ one, a faster switching time can be attained. Once the switching has completed and the error signal approaches zero the $\HV$ part of the circuit becomes inactive again and the $\LV$ power supplies take over. While the load is being supplied by the capacitors their voltage decreases and they become less effective. For this reason this type of design is only useful if the application is compatible with a down time that allows the capacitors to charge.

\begin{figure}
    \centering
    \includegraphics[width=\linewidth]{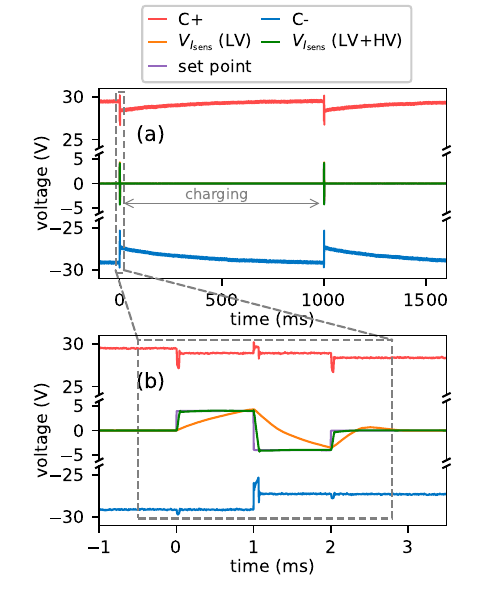}
    \caption{\textbf{Typical operation of the circuit.}  The behavior of the voltage in the positive ($\Cp$) and negative ($\Cm$) capacitors, and current in the inductive load when the current setpoint is changed. The current, presented as the voltage measured at $\VIsens$, is plotted for two different cases: when only the low-voltage supply is used ($\LV$) and when the low and high-voltage supply are used in combination ($\LV+\HV$). (a) Long-term behavior of the circuit. A short pulse sequence is executed at $\mathrm{time}=0$ which depletes the capacitors and is followed by a \SI{1}{\second} wait to allow the capacitors to charge before executing the pulse sequence again. (b) Detailed view of the execution of the time sequence over a short interval around switching events. The charging voltage of the capacitors is \SI{\pm 30}{\volt} and the scale of the current sensor that measures $\VIsens$ is \Iscale. }
    \label{fig:overview}
\end{figure}

\section{Test Setup}
The behavior of the circuit in a typical application is shown in Fig.~\ref{fig:overview}. Here, a short (\SI{2}{\milli\second}) pulse sequence is executed once per second and this delay between pulses allows the capacitors to charge. For the measurements shown in this paper, we used a coil with $L=\SI{491}{\micro\henry}$ and $R=\SI{0.26}{\ohm}$ resulting in a natural time constant $\tau=L/R=\SI{1.9}{\milli\second}$. The $\HV$ stage is powered by two DC power supplies, a GW Instek GPE-3323, whose two $\SI{30}{\volt}$/$\SI{3}{\ampere}$ channels are internally connected in series, and a Rigol DP712. Together, they supply the $\pm \SI{18}{\volt} - \pm \SI{50}{\volt}$ used to charge the KEMET $\SI{1500}{\mu F}$ / $\SI{200}{V}$ capacitors. The $\LV$ stage uses a Rigol DP832 configured in bipolar mode to supply a $\pm \SI{1}{\ampere}$ current to the coil during testing, with a maximum available current of $\pm \SI{3}{\ampere}$ set by the current limit of the power supply. As can be seen in Fig.~\ref{fig:overview}a, the capacitor voltages $\Cp$ and $\Cm$ asymptotically approach the $\HV$ voltage used of \SI{\pm 30}{\volt} during the charging period and drop when the pulse sequence is executed at \SI{0}{\milli\second} and at \SI{1000}{\milli\second}. A more detailed view of the pulse sequence during a short interval around $t=0$ can be observed in Fig.~\ref{fig:overview}b. Here, a comparison is made between the circuit operating only with the $\LV$ part active and when both the $\LV$ and $\HV$ parts are used. When using only the $\LV$ part the value of the current can not even settle before the setpoint has already changed while the case with both $\LV$ and $\HV$ responds much faster and can follow the setpoint better. In the detailed view of the capacitor traces $\Cp$ and $\Cm$ in Fig.~\ref{fig:overview}b it can be seen that when the setpoint jumps upwards the charge of the positive capacitor is used to drive the current and it is, therefore, left with a lower charge after the jump. For downward jumps the negative capacitor is depleted.  Additional to the changes in the $\Cp$ and $\Cm$ voltages due to the decrease in their charge, there are some transient ($\lesssim\SI{0.1}{\milli\second}$) jumps that occur during the activation of the $\HV$ part of the circuit which do not meaningfully affect the capacitor charge since the value before and after the transients is the same when the capacitors charge is not used to supply the current.

\section{Time-domain response}
\begin{figure}
    \centering
    \includegraphics[width=\linewidth]{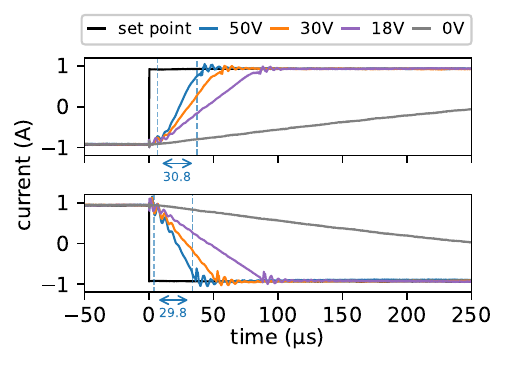}
    \caption{\textbf{Step response of the current.} Response of the current for a rising (top) and falling (bottom) edge on the setpoint for different voltages in the $\HV$ supply. The \SI{0}{\volt} indicates the case where only the $\LV$ part of the circuit is used. For the fastest case, \SI{50}{\volt}, the coil reaches from $10 \%$ to $90 \%$ in $\SI{30.8}{\micro\second}$ on the rising edge and \SI{29.8}{\micro\second} for the falling edge. Using Eq. \ref{eq:CurrentSpeed} the shortest achievable rise or fall time with \SI{50}{\volt} is $\sim \SI{16}{\micro\second}$.}
    \label{fig:time_step}
\end{figure}
To further investigate the switching behavior of the circuit we measure the step behavior for a rising edge and a falling edge for different voltages used for the $\HV$ supply. The current was switched from \SI{-1}{\ampere} to \SI{1}{\ampere} and the obtained results are presented in Fig.~\ref{fig:time_step}. The test with the $\HV$ part disabled is indicated as \SI{0}{\volt}, in this case it takes \SI{900}{\micro\second} for the current to reach the setpoint and further \SI{400}{\micro\second} for it to settle (extending beyond the figure). With the $\HV$ circuit enabled, the switching occurs at a much shorter time scale and increasing the capacitor charge voltage, further reduces the switching time.

\begin{figure}
    \centering
    \includegraphics[width=\linewidth]{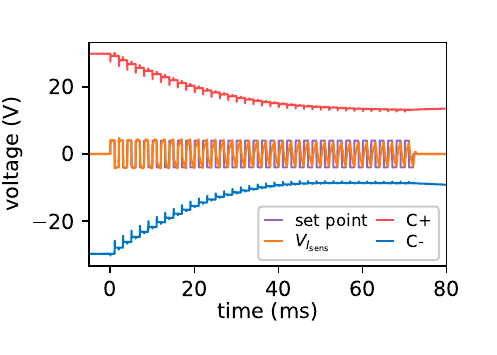}
    \caption{\textbf{Limitations of the fast switching circuit.} A sequence of many square pulses is used as setpoint for the circuit. A stepwise discharge of the capacitors is observed. As the capacitor voltage decreases, the effectiveness of the circuit is reduced. The scale of the current sensor that measures $\VIsens$ is \Iscale. }
    \label{fig:limitations}
\end{figure}

While using higher voltages for switching the current can greatly improve the switching time, this cannot be used for arbitrarily complex pulse sequences due to the finite amount of charge stored in the capacitors. This limitation is demonstrated in Fig.~\ref{fig:limitations}, where a long sequence of square pulses is presented. With decreased capacitor voltage the speed-up in switching time also decreases until there is no improvement at all. Moreover, even though the pulse sequence used in this measurement was symmetrical with respect to \SI{0}{\volt} the remaining charge in the capacitors at the end of the sequence is different. This may be caused by specific differences between the N- and P-channel MOSFETs used in the $\HV$ part, such as differences in resistance when switched completely on ($R_\mathrm{\rs DS(on)}$) or in their gate capacitance.

\section{Frequency-domain response}
\begin{figure}
    \centering
    \includegraphics[width=\linewidth]{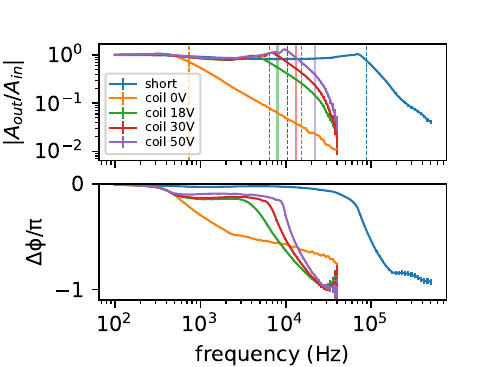}
    \caption{\textbf{Short-term response function.} A sinusoidal input was used as the setpoint of the circuit and the output gain and phase-shift of the first two periods was measured. The ``short'' plot indicates the case where a short wire was used as the load instead of a coil with significant inductance. The ``coil \SI{0}{\volt}'' shows the case with the coil and only the $\LV$ part of the circuit active. The following measurements ``coil \SI{18}{\volt}'', ``coil \SI{30}{\volt}'' and ``coil \SI{50}{\volt}'' show when both the $\LV$ and $\HV$ parts of the circuit are active and the indicated voltage is the one used to charge the capacitors. Vertical dashed lines indicate the bandwidth cutoff frequencies, where the signal magnitude decreases by \SI{3}{\decibel}. The obtained cutoff frequencies are \SI{89.5}{\kilo\hertz}, \SI{0.736}{\kilo\hertz}, \SI{6.5}{\kilo\hertz}, \SI{10.5}{\kilo\hertz}, \SI{15.2}{\kilo\hertz} for the ``short'', ``coil \SI{0}{\volt}'', ``coil \SI{18}{\volt}'', ``coil \SI{30}{\volt}'' and ``coil \SI{50}{\volt}'' respectively. Vertical solid lines are the estimated cutoff frequencies obtained from calculating maximum achievable rise times using Eq. \ref{eq:CurrentSpeed}. The estimated cutoff frequencies are \SI{8.0}{\kilo\hertz}, \SI{13.4}{\kilo\hertz}, \SI{22.3}{\kilo\hertz} for the ``coil \SI{18}{\volt}'', ``coil \SI{30}{\volt}'' and ``coil \SI{50}{\volt}.}
    \label{fig:response_function}
\end{figure}

To gain further understanding of this circuit we also studied it in the frequency domain. A sinusoidal signal was used as the input for the setpoint of the current $\VIset$ and the resulting oscillating current in the load was measured using the current sensor in the circuit to obtain $\VIsens$. Then, for each frequency $f$ of the input, both the input and output signals were fit to $A\sin(2\pi ft-\phi)+ B$ to get the gain $|A_{\rs \mathrm{out}}/A_{\rs \mathrm{in}}|$ and phase-shift $\Delta\phi=\phi_{\rs \mathrm{in}}-\phi_{\rs \mathrm{out}}$ to obtain the response function~\cite{bechhoeferFeedbackPhysicistsTutorial2005}. For this measurement the system was allowed to reach a steady-state at a constant setpoint before sending the oscillating stimulus so that the $\HV$ capacitors started the sequence fully charged. During the analysis we only used the first two sinusoidal periods of the signals to avoid the capacitors' depletion affecting the measurements.  The results are presented in Fig.~\ref{fig:response_function}. To quantify the bandwidth limitation of the control electronics we first measured the response of the circuit for the case where the load was purely resistive and of low resistance avoiding the delays caused by an inductive load. The results of this case are labeled as ``short'' and a bandwidth of $\SI{89.5}{\kilo\hertz}$ was obtained. Once the inductive load was introduced, we first measured with only the $\LV$ part of the circuit active (shown as ``coil \SI{0}{\volt}'') obtaining a bandwidth of $\SI{736}{\hertz}$. With the $\HV$ part of the circuit enabled  we observe an increase in bandwidth by more than a factor of $20$. Again, with higher $\HV$ voltage there is a larger increase in bandwidth. This is consistent with step measurements of Fig.~\ref{fig:time_step} since the more frequency components available to reproduce a step function, the shorter the rising and falling times will be. At low frequency, the response of all the cases with inductive load is identical since the $\HV$ part of the circuit does not activate. To activate the $\HV$ part it is required that the error signal overcomes a threshold and  for slow signals, the $\LV$ part is fast enough to follow the setpoint. Only when the $\LV$ part of the circuit begins to falter the $\HV$ activates.

To understand why the bandwidth increases when activating the $\HV$ section, the dynamical system~\cite{ogataModernControlEngineering2010} represented by the block diagram shown in Fig.~\ref{fig:circuit}b is a good approach for analysis. The shaded region shows that the error signal gets amplified by a factor $g$, then there is a dead-zone block that models the voltage threshold $\VGSth$ required by each $\HV$ MOSFET to start conducting. Finally another gain stage by a factor $h$ models both the $\HV$ MOSFETs and the charged capacitors.  The amplified signal is then added to the output of the PID to drive the load. It is important to note that this model is only valid for short pulse bursts, since for longer times the capacitors discharge effectively decreasing the $h$ gain as in Fig.~\ref{fig:limitations}. The overall effect of the shaded part of the circuit is that, whenever the error signal overcomes the threshold set by the dead-zone, the proportional gain of the feedback circuit is increased to $\approx K_p + gh$, where $K_p$ is the proportional gain of the PID in the $\LV$ part of the circuit. When the gain inside a control loop is increased, the system  can follow the setpoint more closely and the bandwidth is effectively increased~\cite{bechhoeferFeedbackPhysicistsTutorial2005}.

\section{Scaling and limits}
A feature of this circuit is that shorter switching times can be achieved by increasing the capacitor charging voltage which could even be in the range of several \si{\kilo\volt} when using a switched-mode power supply such as a DC-DC converter. However, there are several practical limitations to using such high voltages which will ultimately constrain scaling this solution: i) For MOSFETs, their resistance when switched completely on, $R_\mathrm{\rs DS(on)}$, grows as the square of the maximum rated voltage that the MOSFET can handle~\cite{horowitzArtElectronics2024}. Therefore, when increasing the voltage rating, at some point the dissipated power might become unmanageable. ii) Capacitor with both a high breakdown voltage and a high capacitance are uncommon so a large array of capacitors may be required. iii) Higher voltage circuits also require special design considerations in the geometry and spacing of the PCB traces as well as conformal coating to prevent insulation breakdown~\cite{IPC2221}.

The selection of the required capacitors is very application dependent. For a use case where only a small jump in current is required every few seconds a small capacitor may suffice while for a complex sequence a larger capacitance may be required. This would slow down the voltage decrease on the capacitors but will also take a longer time to charge them for the same output current of the $\HV$ supply.

When designing an electromagnet to achieve a specific magnetic field $B_0$ in a given geometry, the product $I_0 N$ of the required current delivered to the coil $I_0$ and the number of windings $N$ is fixed by $B_0$. A coil with many windings and a small current would produce the same field as one with few windings and a large current. For experiments where a fast switching of the coils is desirable, the number of windings is minimized at the expense of increased current requirements, and increased wire thickness, to decrease the inductance of the coil. In the context of this circuit the choice of fewer windings has the additional advantage that the voltage drop at the coil in the steady-state operation (the $\LV$ part) is small. If, instead, the coil required large voltages even for the steady-state operation, then the $\HV$ part of the circuit would not provide such a great advantage. Additionally, the maximum voltage that can be applied to the coil is bounded in the presented implementation by the voltage range of the PID control electronics to drive the $\LV$ MOSFETs. Since many magnetic coils designed for cold atom experiments are already optimized for fast on-off switching by minimizing inductance, they are well-suited for applications using this circuit.

\section{Comparison with other devices}
Multiple strategies are available for controlling the current flowing through magnetic coils. Some designs, include the BEC Coil Driver from Stummer~\cite{stummer2007} which relies on a finely tuned pulse from a large storage capacitor for switching on, and a string of transient-voltage-suppression (TVS) diodes for fast shutdown. The circuit developed by Claussen~\cite{ClaussenDynamicsBoseCondensate2003} similarly uses a capacitor bank, but with added control for unipolar current modulation. In contrast, our on-demand capacitor boost design supports both dynamic and bipolar current control. 

Designs based on high-power operational amplifiers offer excellent precision~\cite{sabulskyCurrentControllersOptimizing2024}. However, while their current can be increased using operational amplifiers in parallel~\cite{YangBipolarCurrentDriver2019} they are inherently less scalable in terms of maximum voltage due to the operational amplifier limitations. In comparison, our implementation can be modified for higher transient voltages to allow faster rising and falling times.

Industrial instruments, such as the Kepco BOP-L, Matsusada DOS or AETechron 7200 series, are appropriate to manage inductive loads and rapid current changes, however are optimized for either high current or high voltage, but not both simultaneously. Instead, our circuit is intended to be an addition to a general purpose power supply. By combining a high current steady-state supply and another enabling high-voltage transients, our circuit can deliver short duration power levels significantly higher than what either supply could achieve individually. Moreover, the principle of the presented circuit could be scaled up for currents in the hundreds of amperes while the mentioned commercial bipolar drivers are limited by the available products. 



We compared our system with the Kepco BOP-L series linear magnet power supplies. According to Eq.~\ref{eq:CurrentSpeed}, the highest voltage model, the BOP 200-1L, would ideally be able to switch our test coil from \SI{-1}{\ampere} to \SI{+1}{\ampere} within \SI{4.9}{\micro\second}, corresponding to a bandwidth of \SI{71}{\kilo\hertz}. However, the bandwidth of the entire BOP-L series is limited to \SI{10}{\kilo\hertz}~\cite{kepcoBOP}. Our driver therefore offers a higher modulation bandwidth. This performance advantage is only transient, as it relies on a capacitor bank whose voltage decays during operation, leading to bandwidth degradation for extended pulse trains, unless the bank is periodically recharged.

\section{Conclusions}

The presented design consists of two pairs of MOSFETs working together in a coordinated manner. One regulates the current from the $\LV$ supply and has the typical performance of circuits driving magnetic coils in cold atom experiments. The second pair, which provides current from capacitors charged to a $\HV$, results in a high improvement to the controller in terms of switching time and bandwidth. Switching time is reduced from $ \SI{900}{\micro\second}$ to $ \SI{30}{\micro\second}$, while bandwidth is improved from $\sim\SI{736}{\hertz}$ when only the $\LV$ circuit is enabled to $\SI{15.2}{\kilo\hertz}$. 

This circuit is a solution to control abrupt current steps including changes in current direction. Moreover, it can follow arbitrary control signals that are within the demonstrated bandwidth and the limitation of cycles needed to discharge the capacitors. It can be adapted to magnetic coils in cold atoms experiments that are normally driven by power supplies which are limited to deliver a large steady-state current operating at low voltages. 

By an appropriate component selection it can be adapted for different needs by changing the capacitors, MOSFETs and tuning the $\HV$ activation threshold. More consecutive current pulses can be achieved with larger capacitors while a faster reset of the circuit's operating state is obtained when using capacitors with lower capacitance.  In other words, the great versatility of our proposal allows it to be adapted to a wide variety of situations, depending on the user's needs.

The demonstrated fast control opens the possibility of implementing improved experimental sequences by allowing more agile progress through the different experimental stages. It can also improve existing cooling and state-preparation stages and allow the exploration of different physical regimes in which the time required for magnetic fields to change must be as short as possible. For example, experiments in which the inter-particle interaction is controlled by a magnetic field via a Fano-Feshbach resonance, can greatly benefit from our scheme, as it allows avoiding delay times and the resulting decrease in the cloud's phase-space density, since the waiting time to reach the desired magnetic field can be significantly reduced.

\section*{Acknowledgements}
The authors would like to acknowledge Alan Stummer for publishing the circuits he developed for magnetic coil switching which served as an inspiration, Michael Schlagmüller whose original PID circuit design we adapted in this circuit, Rodrigo Alejandro Gutiérrez Arenas for early discussions on conceptualizing this work and Rebeca Díaz-Pardo for test equipment. This work was supported by the UNAM-CIC LANMAC program. J. A. S. acknowledges support from UNAM-PAPIIT IN105724 and SECIHTI CF-2023-I-72 and A. P-M. acknowledges support from UNAM-PAPIIT IN115523. L. U-R, E. G. A-T., A. H-L., E. E-R, and G. C-T., acknowledge fellowships from SECIHTI for their graduate studies.

\appendix*

\section{Full Schematic}
\label{sec:FullSchematic}
\begin{figure*}
    \centering
    \includegraphics[width=\linewidth]{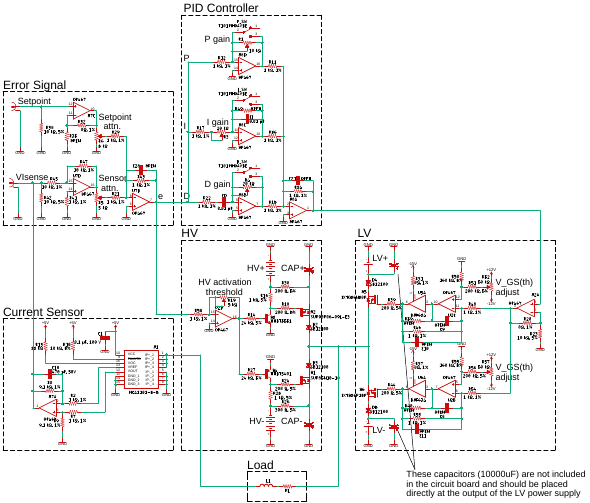}
    \caption{\textbf{Full schematic of the circuit.} The work files and PCB layout can be obtained from the repository~\cite{repo}.}
    \label{fig:full_schematic}
\end{figure*}

The full schematic is shown in Fig.~\ref{fig:full_schematic} and the work files can be obtained from the repository\cite{repo}. The schematic is divided in several sections and their function and the design choices made in each case will be described following these categories.

\subsection{Current Sensor}
The current sensor is a critical part of the circuit as its precision and noise impacts the performance of the whole circuit. In our case, the current passing through the load is measured by the \texttt{MCA1101-5-5} sensor. It works by magnetic means and does not cause a voltage drop, as opposed to using small resistor in series with the load. This sensor is rated for $\pm\SI{5}{\ampere}$ but there are also $\pm\SI{20}{\ampere}$ and $\pm\SI{50}{\ampere}$ varieties that could be used in case a wider range is required.

The amplifier \texttt{U7A} after the current sensor is there to adapt the output scale of the current sensor to the operating voltage range of the circuit to minimize the impact of noise. Additionally, a capacitor was added to its feedback loop to filter out noise from the sensor.


\subsection{Error Signal}
This part of the circuit has an input $\VIset$ for setting the current value and an output $\VIsens$ to read the measurement of the current sensor. Each of these voltages has an independent attenuator so that the voltage scale of each channel can be adjusted to suit any control system or signal generator. As the \texttt{U7D} amplifier used for $\VIsens$ is set on an inverting configuration, adding both attenuated signals on \texttt{U7B} results in the error signal $e=\VIset-\VIsens$ which is fed into the PID controller.

\subsection{PID Controller}
The PID controller is laid out using three separate operational amplifiers that are added by the \texttt{U8A} amplifier. As opposed to a single opamp PID design, this arrangement simplifies tuning of the P, I, and D parameters independently by adjusting the \texttt{R1}, \texttt{R3} and \texttt{R4} trimpots respectively. Additionally, the different contributions of the controller can be disabled with the \texttt{P\_SW}, \texttt{I\_SW}, \texttt{D\_SW} switches. For most applications and in all of our testing the D component of the controller did not provide any advantage.

\subsection{Low-voltage}
The PID output is fed into the $\LV$ section of the circuit by the \texttt{U2A} amplifier that acts as a buffer. Its output is then split to drive the positive and negative MOSFET drivers consisting of the \texttt{U2C} and \texttt{U5A} for the positive part and  \texttt{U2B} and \texttt{U6A} for the negative one. The positive driver works by using \texttt{U2C} to add an offset, set by \texttt{R52}, to the output of the PID and then using the \texttt{U5A} buffer to provide the high current required to quickly drive the MOSFET gate. The offset is added to cancel out the gate-source threshold voltage $\VGSth$ so that the current in the coils can transition smoothly between negative and positive values. The negative MOSFET driver works analogously. Commercial integrated MOSFET drivers are usually designed for on-off applications and therefore we opted for a discrete solution to be able to generate other types of waveforms.

The supply powering the $\LV$ part of the circuit should be operated in constant voltage mode. To help it maintain a constant voltage even for a fast varying load, capacitors of $\SI{10000}{\micro\farad}$ are placed directly at its positive and negative outputs. 

The MOSFETs, \texttt{Q5} and \texttt{Q6}, are the only power components of this section that must be able to handle the large currents passing through the magnetic coils. In order to scale the circuit for larger currents, several MOSFETs could be connected in parallel or alternative components designed for larger currents can be used.

\subsection{High-voltage}
The high-voltage section of the circuit is fed directly with the error signal. It is then scaled up with \texttt{U2D} by a factor adjusted with \texttt{R19} and used to drive \texttt{Q3} and \texttt{Q4} transistors. As the threshold voltage of the transistors is not compensated they only begin conduction when the output of \texttt{U2D} has reached these voltages. Therefore, adjusting the scaling factor changes the value of the error needed to activate the transistors. 

Having a threshold-based mechanism allows for the $\HV$ capacitors to discharge only when necessary. If a current ramp is desired, slow enough for the $\LV$ section to be able to follow, the error signal stays small and the $\HV$ circuit does not activate. As soon as the ramp becomes too fast for the $\LV$ part to follow, the error will quickly grow and activate the $\HV$ section helping the current to catch up to the setpoint value. 

In the $\HV$ section the \texttt{Q1} and \texttt{Q2} MOSFETs are arranged in the opposite way as in the $\LV$ part of the circuit: The P-channel MOSFET is connected to the positive supply while the N-channel one is connected to the negative supply. This means that by themselves they do not act as push-pull arrangement. However, by tying their gates to the $\HVp$ and $\HVm$ supplies using \texttt{R16} and \texttt{R25} respectively, the \texttt{Q3} and \texttt{Q4} transistors can control the flow of current through the MOSFETs obtaining an effective push-pull configuration. In this case, the gate voltage of the MOSFETs is referenced to the $\HVp$ and $\HVm$ supplies and not to the voltage across the load as in the $\LV$ section. The advantage of using this arrangement is that the current through the load can be controlled even if the load voltage is above the supply voltage of the opamps. 

To better understand how the $\HV$ threshold works, we can focus on the positive part of the circuit. When the transistor \texttt{Q3} is not conducting, the gate voltage of the \texttt{Q2} MOSFET is pulled up to the $\HVp$ supply. Since it is a P-channel MOSFET there is no conduction in this configuration. Once the \texttt{Q3} transistor begins conducting, the current through \texttt{R16} results in a voltage drop that opens the gate of \texttt{Q2}. With this logic we can determine the voltage at the output of the \texttt{UD2} opamp required to activate the \texttt{Q2} MOSFET by working backwards starting from the MOSFET. This voltage is given by
\[
V_\mathrm{th}^{\HVp} = \frac{R_{14} V_\mathrm{GS(th)}^\texttt{Q2}}{R_{16} h_\mathrm{FE}^\texttt{Q3}} + V_\mathrm{BE}^\texttt{Q3},
\]
where $R_{14}$ and $R_{16}$ are the resistance values of \texttt{R14} and \texttt{R16}, $V_\mathrm{GS(th)}^\texttt{Q2}$ is the gate-source threshold voltage of \texttt{Q2} and $V_\mathrm{BE}^\texttt{Q3}$ is the base-emitter voltage of \texttt{Q3}.

Analogously, for the negative part we get
\[
V_\mathrm{th}^{\HVm} = \frac{R_{27} V_\mathrm{GS(th)}^\texttt{Q1}}{R_{25} h_\mathrm{FE}^\texttt{Q4}} + V_\mathrm{BE}^\texttt{Q4},
\]
following the same naming scheme.

\subsubsection{Safety considerations}
Given the flexibility of this design and the high-voltages that can be present in the circuit some safety measures should be considered. For example, an interlock can be included to disable the $\HV$ supply in case the enclosure of the circuit is opened. Another measure is to have bleeder resistors at the $\HV$ capacitors that will cause them to discharge when the circuit is powered off. This resistor should be chosen such that the capacitor discharge rate should be considerably slower than the charging rate.

\subsection{Load}
The maximum load that can be controlled with this circuit is limited by the architecture of the $\LV$ section. Since the MOSFETs are driven by opamps, the maximum voltage that can be applied to the gate of the $\LV$ MOSFETs is $\pm\SI{13}{\volt}$ for the \texttt{OP467} opamps supplied at $\pm\SI{15}{\volt}$. Moreover the $\VGSth$ can range from \SI{3}{\volt} to \SI{5.5}{\volt} for the \texttt{IXTQ36N30P} MOSFET and since it is referenced by the high-side of the load, the maximum load voltage is between \SI{10}{\volt} and \SI{7.5}{\volt}. Analogously, for the negative part of the circuit the load voltage is between \SI{-10.5}{\volt} and \SI{-8.5}{\volt}.

This imposes an important limit on the size of coils that can be driven. For a \SI{1}{\ohm} coil the maximum steady-state current is \SI{7.5}{\ampere} in the worst case scenario for the MOSFET's $\VGSth$.

\subsection{Supplies}
In addition to the $\LV$ and $\HV$ power supplies, the circuit requires a $\pm\SI{15}{\volt}$ supply to power the opamps and analog electronics. The circuit includes regulators to obtain \SI{5}{\volt} for the current sensor and $\pm\SI{12}{\volt}$ to have a stable reference for the \texttt{R52} and \texttt{R57} resistors used to set the offset voltages of the $\LV$ section.

\subsection{First Adjustment}
After assembling the circuit it first needs to be adjusted to the load and to account for component variations for it to operate correctly. 
\subsubsection{Offset compensation}
The offsets added to the output of the PID signal must be set in order to have a smooth crossover from negative to positive currents. To do so we adjust them until a small current of \SI{20}{\milli\ampere} flows through the \texttt{Q5} and \texttt{Q6} MOSFETs even at zero setpoint. Since the same current flows through both MOSFETs this constant bias current is not diverted to the load and still ensures a smooth transition. The following procedure is used:
\begin{enumerate}
\itemsep0em 
    \item Connect the coil to be used to the circuit.
    \item With the $\HV$ and $\LV$ supplies disabled, set \SI{0}{\volt} as the setpoint and disable the PID using the \texttt{P\_SW}, \texttt{I\_SW}, \texttt{D\_SW} switches.
    \item Turn on the $\LVp$ supply and adjust \texttt{R52} until the circuit draws \SI{20}{\milli\ampere} for the supply.
    \item Turn off $\LVp$ and repeat the previous step with $\LVm$ and \texttt{R57}.
\end{enumerate}

\subsubsection{PID}
The PID controller is only used by the $\LV$ part of the circuit and it is best to first adjust it with the $\HV$ supply off. Otherwise, a poor adjustment of these parameters could result in a large error signal that would constantly activate the $\HV$ part. Once the PID has been optimized using only the $\LV$ supply it can then be fine tuned with both the $\LV$ and $\HV$ supplies enabled.

\subsubsection{Setting the $\HV$ threshold}
To adjust the $\HV$ threshold a \SI{1}{\hertz} square wave was used for the setpoint that would simulate a typical desired current jump. The activation threshold is then slowly increased by adjusting \texttt{R19} until the $\HV$ part activates when desired. 

\section{Thermal Considerations}
The components that handle the most power in the circuit are the MOSFETs that drive current into the load. However, as the $\HV$ MOSFETs are only active during very short transients their heating was not an issue. Therefore, a thermal analysis was carried out to make sure only the $\LV$ regulating MOSFETs, \texttt{IXTQ36P15P} and \texttt{IXTQ36N30P}, operate below their junction temperature limit $T_{\rs\mathrm{J, max}} = \SI{150}{\celsius}$. We calculate this temperature assuming the circuit is powered by the maximum available supply current of $\SI{3}{\ampere}$.

The MOSFET's thermal path consists of three successive stages: junction-to-case, case-to-heatsink and heatsink-to-ambient, and their dissipated power is given by
\begin{equation}
    P_{\rs\mathrm{MOSFET}} = \frac{T_{\rs\mathrm{J}}-T_{\rs\mathrm{A}}}{R_{\rs\mathrm{\theta JC}} + R_{\rs\mathrm{\theta\text{CS}}} + R_{\rs\mathrm{\theta\text{SA}}}} 
    \label{eq:apdx_power_mosfets_temperature}
\end{equation}

\noindent
where $T_{\rs\mathrm{J}}$ is the junction temperature of the MOSFET, $T_{\rs\mathrm{A}}$ is the ambient temperature and the terms, $R_{\rs\mathrm{\theta {JC}}}$, $R_{\rs\mathrm{\theta {CS}}}$ and $R_{\rs\mathrm{\theta SA}}$, are the thermal resistances of the junction-to-case, case-to-heatsink and heatsink-to-ambient, respectively. 

The highest power dissipation in the \texttt{IXTQ36P15P} occurs when the MOSFET's drain-source resistance $R_{\rs\mathrm{DS}}$ matches with the coil's resistance $R_{\rs\mathrm{L}}$. In this condition, half of the applied power is dropped across the load and half across the MOSFET. Having a drain current of $I_{\rs\mathrm{D}}=\SI{3}{\ampere}$, the power dissipated by the MOSFET is,
\begin{equation}
  P_{\rs\mathrm{MOSFET}} = I_{\rs\mathrm{D}}^{2} R_{\rs\mathrm{DS}} = I_{\rs\mathrm{D}}^{2} R_{\rs\mathrm{L}} = \SI{2.34}{\watt} .
  \label{eq:apdx_mosfet_power}
\end{equation}

Based on the manufacturer’s data and the specifications of the selected Wakefield Engineering heatsink model 403K, we obtain $R_{\rs\mathrm{\theta {JC}}} = \SI{0.42}{\celsius\watt^{-1}}$, $R_{\rs\mathrm{\theta CS}} = \SI{0.21}{\celsius\watt^{-1}}$, $R_{\rs\mathrm{\theta SA}} = \SI{1.83}{\celsius\watt^{-1}}$. Summing these values gives a total thermal resistances of $R_{\rs\mathrm{\theta tot}} = \SI{2.46}{\celsius\watt^{-1}}$. Finally, combining Eqs. \ref{eq:apdx_power_mosfets_temperature} and \ref{eq:apdx_mosfet_power} gives an estimated junction temperature of 
\begin{equation}
\begin{split}
    T_{\rs\mathrm{J}} &= P_{\rs\mathrm{MOSFET}} R_{\rs\mathrm{\theta tot}} + T_{\rs\mathrm{A}} = \SI{25.7}{\celsius} ,
\end{split}
\end{equation}

\noindent
where $T_{\rs\mathrm{A}} = \SI{20}{\celsius}$. Applying the same procedure to the \texttt{IXTQ36N30P} yields the same result, as its electrical and thermal parameters are identical to those of the \texttt{IXTQ36P15P}. The resulting junction-temperature is again $\SI{25.7}{\celsius}$ for this model.

Both MOSFETs thus operate within safe conditions, considering that the maximum junction temperature specified in their data-sheets is $T_{\rs\mathrm{J, max}} = \SI{150}{\celsius}$.

We validated the thermal model by performing direct temperature measurements on the MOSFETs under load after several minutes of operation at the maximum $\pm\SI{3}{\ampere}$. The highest temperature recorded was about \SI{23.5}{\celsius}.




Although the calculated and measured temperatures are well below $T_{\rs\mathrm{J, max}}$, higher operating currents leads to greater power dissipation requirements. In such situations, this design allows the use of multiple MOSFETs in parallel, so the electrical and thermal load can be divided among several transistors. 


\bibliographystyle{apsrev4-2}
\bibliography{bibliography}

\begin{thebibliography}{33}%
\makeatletter
\providecommand \@ifxundefined [1]{%
 \@ifx{#1\undefined}
}%
\providecommand \@ifnum [1]{%
 \ifnum #1\expandafter \@firstoftwo
 \else \expandafter \@secondoftwo
 \fi
}%
\providecommand \@ifx [1]{%
 \ifx #1\expandafter \@firstoftwo
 \else \expandafter \@secondoftwo
 \fi
}%
\providecommand \natexlab [1]{#1}%
\providecommand \enquote  [1]{``#1''}%
\providecommand \bibnamefont  [1]{#1}%
\providecommand \bibfnamefont [1]{#1}%
\providecommand \citenamefont [1]{#1}%
\providecommand \href@noop [0]{\@secondoftwo}%
\providecommand \href [0]{\begingroup \@sanitize@url \@href}%
\providecommand \@href[1]{\@@startlink{#1}\@@href}%
\providecommand \@@href[1]{\endgroup#1\@@endlink}%
\providecommand \@sanitize@url [0]{\catcode `\\12\catcode `\$12\catcode
  `\&12\catcode `\#12\catcode `\^12\catcode `\_12\catcode `\%12\relax}%
\providecommand \@@startlink[1]{}%
\providecommand \@@endlink[0]{}%
\providecommand \url  [0]{\begingroup\@sanitize@url \@url }%
\providecommand \@url [1]{\endgroup\@href {#1}{\urlprefix }}%
\providecommand \urlprefix  [0]{URL }%
\providecommand \Eprint [0]{\href }%
\providecommand \doibase [0]{https://doi.org/}%
\providecommand \selectlanguage [0]{\@gobble}%
\providecommand \bibinfo  [0]{\@secondoftwo}%
\providecommand \bibfield  [0]{\@secondoftwo}%
\providecommand \translation [1]{[#1]}%
\providecommand \BibitemOpen [0]{}%
\providecommand \bibitemStop [0]{}%
\providecommand \bibitemNoStop [0]{.\EOS\space}%
\providecommand \EOS [0]{\spacefactor3000\relax}%
\providecommand \BibitemShut  [1]{\csname bibitem#1\endcsname}%
\let\auto@bib@innerbib\@empty
\bibitem [{\citenamefont {Bloch}\ \emph {et~al.}(2008)\citenamefont {Bloch},
  \citenamefont {Dalibard},\ and\ \citenamefont
  {Zwerger}}]{blochUltracoldGases2008}%
  \BibitemOpen
  \bibfield  {author} {\bibinfo {author} {\bibfnamefont {I.}~\bibnamefont
  {Bloch}}, \bibinfo {author} {\bibfnamefont {J.}~\bibnamefont {Dalibard}},\
  and\ \bibinfo {author} {\bibfnamefont {W.}~\bibnamefont {Zwerger}},\ }\href
  {https://doi.org/10.1103/RevModPhys.80.885} {\bibfield  {journal} {\bibinfo
  {journal} {Rev. Mod. Phys.}\ }\textbf {\bibinfo {volume} {80}},\ \bibinfo
  {pages} {885} (\bibinfo {year} {2008})}\BibitemShut {NoStop}%
\bibitem [{\citenamefont {Schlosser}\ \emph {et~al.}(2001)\citenamefont
  {Schlosser}, \citenamefont {Reymond}, \citenamefont {Protsenko},\ and\
  \citenamefont {Grangier}}]{schlosserLoadingSingleAtoms2001}%
  \BibitemOpen
  \bibfield  {author} {\bibinfo {author} {\bibfnamefont {N.}~\bibnamefont
  {Schlosser}}, \bibinfo {author} {\bibfnamefont {G.}~\bibnamefont {Reymond}},
  \bibinfo {author} {\bibfnamefont {I.}~\bibnamefont {Protsenko}},\ and\
  \bibinfo {author} {\bibfnamefont {P.}~\bibnamefont {Grangier}},\ }\href
  {https://doi.org/10.1038/35082512} {\bibfield  {journal} {\bibinfo  {journal}
  {Nature}\ }\textbf {\bibinfo {volume} {411}},\ \bibinfo {pages} {1024}
  (\bibinfo {year} {2001})}\BibitemShut {NoStop}%
\bibitem [{\citenamefont {Kaufman}\ and\ \citenamefont
  {Ni}(2021)}]{kaufmanTwizzerArraysUltracoldAtoms2021}%
  \BibitemOpen
  \bibfield  {author} {\bibinfo {author} {\bibfnamefont {A.~M.}\ \bibnamefont
  {Kaufman}}\ and\ \bibinfo {author} {\bibfnamefont {K.-K.}\ \bibnamefont
  {Ni}},\ }\href {https://doi.org/10.1038/s41567-021-01357-2} {\bibfield
  {journal} {\bibinfo  {journal} {Nat. Phys.}\ }\textbf {\bibinfo {volume}
  {17}},\ \bibinfo {pages} {1324} (\bibinfo {year} {2021})}\BibitemShut
  {NoStop}%
\bibitem [{\citenamefont {Sherson}\ \emph {et~al.}(2010)\citenamefont
  {Sherson}, \citenamefont {Weitenberg}, \citenamefont {Endres}, \citenamefont
  {Cheneau}, \citenamefont {Bloch},\ and\ \citenamefont
  {Kuhr}}]{shersonSingleAtomMottInsulator2010}%
  \BibitemOpen
  \bibfield  {author} {\bibinfo {author} {\bibfnamefont {J.~F.}\ \bibnamefont
  {Sherson}}, \bibinfo {author} {\bibfnamefont {C.}~\bibnamefont {Weitenberg}},
  \bibinfo {author} {\bibfnamefont {M.}~\bibnamefont {Endres}}, \bibinfo
  {author} {\bibfnamefont {M.}~\bibnamefont {Cheneau}}, \bibinfo {author}
  {\bibfnamefont {I.}~\bibnamefont {Bloch}},\ and\ \bibinfo {author}
  {\bibfnamefont {S.}~\bibnamefont {Kuhr}},\ }\href
  {https://doi.org/10.1038/nature09378} {\bibfield  {journal} {\bibinfo
  {journal} {Nature}\ }\textbf {\bibinfo {volume} {467}},\ \bibinfo {pages}
  {68–72} (\bibinfo {year} {2010})}\BibitemShut {NoStop}%
\bibitem [{\citenamefont {Neuhauser}\ \emph {et~al.}(1980)\citenamefont
  {Neuhauser}, \citenamefont {Hohenstatt}, \citenamefont {Toschek},\ and\
  \citenamefont {Dehmelt}}]{neuhauserFirstSingleIon1980}%
  \BibitemOpen
  \bibfield  {author} {\bibinfo {author} {\bibfnamefont {W.}~\bibnamefont
  {Neuhauser}}, \bibinfo {author} {\bibfnamefont {M.}~\bibnamefont
  {Hohenstatt}}, \bibinfo {author} {\bibfnamefont {P.~E.}\ \bibnamefont
  {Toschek}},\ and\ \bibinfo {author} {\bibfnamefont {H.}~\bibnamefont
  {Dehmelt}},\ }\href {https://doi.org/10.1103/PhysRevA.22.1137} {\bibfield
  {journal} {\bibinfo  {journal} {Phys. Rev. A}\ }\textbf {\bibinfo {volume}
  {22}},\ \bibinfo {pages} {1137} (\bibinfo {year} {1980})}\BibitemShut
  {NoStop}%
\bibitem [{\citenamefont {Wineland}\ and\ \citenamefont
  {Itano}(1981)}]{winelandSingleMgIon1981}%
  \BibitemOpen
  \bibfield  {author} {\bibinfo {author} {\bibfnamefont {D.}~\bibnamefont
  {Wineland}}\ and\ \bibinfo {author} {\bibfnamefont {W.~M.}\ \bibnamefont
  {Itano}},\ }\href {https://doi.org/10.1016/0375-9601(81)90942-7} {\bibfield
  {journal} {\bibinfo  {journal} {Phys. Lett. A}\ }\textbf {\bibinfo {volume}
  {82}},\ \bibinfo {pages} {75–78} (\bibinfo {year} {1981})}\BibitemShut
  {NoStop}%
\bibitem [{\citenamefont {Zipkes}\ \emph {et~al.}(2010)\citenamefont {Zipkes},
  \citenamefont {Palzer}, \citenamefont {Sias},\ and\ \citenamefont
  {K\"{o}hl}}]{zipkesSingleIonBoseCondensate2010}%
  \BibitemOpen
  \bibfield  {author} {\bibinfo {author} {\bibfnamefont {C.}~\bibnamefont
  {Zipkes}}, \bibinfo {author} {\bibfnamefont {S.}~\bibnamefont {Palzer}},
  \bibinfo {author} {\bibfnamefont {C.}~\bibnamefont {Sias}},\ and\ \bibinfo
  {author} {\bibfnamefont {M.}~\bibnamefont {K\"{o}hl}},\ }\href
  {https://doi.org/10.1038/nature08865} {\bibfield  {journal} {\bibinfo
  {journal} {Nature}\ }\textbf {\bibinfo {volume} {464}},\ \bibinfo {pages}
  {388–391} (\bibinfo {year} {2010})}\BibitemShut {NoStop}%
\bibitem [{\citenamefont {Chang}\ \emph {et~al.}(2014)\citenamefont {Chang},
  \citenamefont {Vuleti{\'c}},\ and\ \citenamefont
  {Lukin}}]{changQuantumNonlinearOptics2014}%
  \BibitemOpen
  \bibfield  {author} {\bibinfo {author} {\bibfnamefont {D.~E.}\ \bibnamefont
  {Chang}}, \bibinfo {author} {\bibfnamefont {V.}~\bibnamefont {Vuleti{\'c}}},\
  and\ \bibinfo {author} {\bibfnamefont {M.~D.}\ \bibnamefont {Lukin}},\ }\href
  {https://doi.org/10.1038/nphoton.2014.192} {\bibfield  {journal} {\bibinfo
  {journal} {Nat. Photonics}\ }\textbf {\bibinfo {volume} {8}},\ \bibinfo
  {pages} {685} (\bibinfo {year} {2014})}\BibitemShut {NoStop}%
\bibitem [{\citenamefont {Metcalf}\ and\ \citenamefont {der
  Straten}(1999)}]{Metcalf1999}%
  \BibitemOpen
  \bibfield  {author} {\bibinfo {author} {\bibfnamefont {H.~J.}\ \bibnamefont
  {Metcalf}}\ and\ \bibinfo {author} {\bibfnamefont {P.~V.}\ \bibnamefont {der
  Straten}},\ }\href {https://link.springer.com/book/10.1007/978-1-4612-1470-0}
  {\emph {\bibinfo {title} {Laser Cooling and Trapping}}}\ (\bibinfo
  {publisher} {Springer},\ \bibinfo {address} {New York},\ \bibinfo {year}
  {1999})\BibitemShut {NoStop}%
\bibitem [{\citenamefont {Bergeman}\ \emph {et~al.}(1987)\citenamefont
  {Bergeman}, \citenamefont {Erez},\ and\ \citenamefont
  {Metcalf}}]{bergemanMagnetostaticTrappingFieldsNeutralAtoms}%
  \BibitemOpen
  \bibfield  {author} {\bibinfo {author} {\bibfnamefont {T.}~\bibnamefont
  {Bergeman}}, \bibinfo {author} {\bibfnamefont {G.}~\bibnamefont {Erez}},\
  and\ \bibinfo {author} {\bibfnamefont {H.~J.}\ \bibnamefont {Metcalf}},\
  }\href {https://doi.org/10.1103/PhysRevA.35.1535} {\bibfield  {journal}
  {\bibinfo  {journal} {Phys. Rev. A}\ }\textbf {\bibinfo {volume} {35}},\
  \bibinfo {pages} {1535} (\bibinfo {year} {1987})}\BibitemShut {NoStop}%
\bibitem [{\citenamefont {Migdall}\ \emph {et~al.}(1985)\citenamefont
  {Migdall}, \citenamefont {Prodan}, \citenamefont {Phillips}, \citenamefont
  {Bergeman},\ and\ \citenamefont
  {Metcalf}}]{phillipsFirstObservationTrappedNeutralAtoms}%
  \BibitemOpen
  \bibfield  {author} {\bibinfo {author} {\bibfnamefont {A.~L.}\ \bibnamefont
  {Migdall}}, \bibinfo {author} {\bibfnamefont {J.~V.}\ \bibnamefont {Prodan}},
  \bibinfo {author} {\bibfnamefont {W.~D.}\ \bibnamefont {Phillips}}, \bibinfo
  {author} {\bibfnamefont {T.~H.}\ \bibnamefont {Bergeman}},\ and\ \bibinfo
  {author} {\bibfnamefont {H.~J.}\ \bibnamefont {Metcalf}},\ }\href
  {https://doi.org/10.1103/PhysRevLett.54.2596} {\bibfield  {journal} {\bibinfo
   {journal} {Phys. Rev. Lett.}\ }\textbf {\bibinfo {volume} {54}},\ \bibinfo
  {pages} {2596} (\bibinfo {year} {1985})}\BibitemShut {NoStop}%
\bibitem [{\citenamefont {Greiner}\ \emph {et~al.}(2001)\citenamefont
  {Greiner}, \citenamefont {Bloch}, \citenamefont {H\"ansch},\ and\
  \citenamefont {Esslinger}}]{esslingerMagneticTransportTrappedColdAtoms}%
  \BibitemOpen
  \bibfield  {author} {\bibinfo {author} {\bibfnamefont {M.}~\bibnamefont
  {Greiner}}, \bibinfo {author} {\bibfnamefont {I.}~\bibnamefont {Bloch}},
  \bibinfo {author} {\bibfnamefont {T.~W.}\ \bibnamefont {H\"ansch}},\ and\
  \bibinfo {author} {\bibfnamefont {T.}~\bibnamefont {Esslinger}},\ }\href
  {https://doi.org/10.1103/PhysRevA.63.031401} {\bibfield  {journal} {\bibinfo
  {journal} {Phys. Rev. A}\ }\textbf {\bibinfo {volume} {63}},\ \bibinfo
  {pages} {031401} (\bibinfo {year} {2001})}\BibitemShut {NoStop}%
\bibitem [{\citenamefont {Happer}(1972)}]{happerOpticalPumping1972}%
  \BibitemOpen
  \bibfield  {author} {\bibinfo {author} {\bibfnamefont {W.}~\bibnamefont
  {Happer}},\ }\href {https://doi.org/10.1103/revmodphys.44.169} {\bibfield
  {journal} {\bibinfo  {journal} {Rev. Mod. Phys.}\ }\textbf {\bibinfo {volume}
  {44}},\ \bibinfo {pages} {169–249} (\bibinfo {year} {1972})}\BibitemShut
  {NoStop}%
\bibitem [{\citenamefont {Chin}\ \emph {et~al.}(2010)\citenamefont {Chin},
  \citenamefont {Grimm}, \citenamefont {Julienne},\ and\ \citenamefont
  {Tiesinga}}]{chengFeshbackResonancesUltracolGases2010}%
  \BibitemOpen
  \bibfield  {author} {\bibinfo {author} {\bibfnamefont {C.}~\bibnamefont
  {Chin}}, \bibinfo {author} {\bibfnamefont {R.}~\bibnamefont {Grimm}},
  \bibinfo {author} {\bibfnamefont {P.}~\bibnamefont {Julienne}},\ and\
  \bibinfo {author} {\bibfnamefont {E.}~\bibnamefont {Tiesinga}},\ }\href
  {https://doi.org/10.1103/RevModPhys.82.1225} {\bibfield  {journal} {\bibinfo
  {journal} {Rev. Mod. Phys.}\ }\textbf {\bibinfo {volume} {82}},\ \bibinfo
  {pages} {1225} (\bibinfo {year} {2010})}\BibitemShut {NoStop}%
\bibitem [{\citenamefont {Grier}\ \emph {et~al.}(2013)\citenamefont {Grier},
  \citenamefont {{Ferrier-Barbut}}, \citenamefont {Rem}, \citenamefont
  {Delehaye}, \citenamefont {Khaykovich}, \citenamefont {Chevy},\ and\
  \citenamefont {Salomon}}]{Grier2013a}%
  \BibitemOpen
  \bibfield  {author} {\bibinfo {author} {\bibfnamefont {A.~T.}\ \bibnamefont
  {Grier}}, \bibinfo {author} {\bibfnamefont {I.}~\bibnamefont
  {{Ferrier-Barbut}}}, \bibinfo {author} {\bibfnamefont {B.~S.}\ \bibnamefont
  {Rem}}, \bibinfo {author} {\bibfnamefont {M.}~\bibnamefont {Delehaye}},
  \bibinfo {author} {\bibfnamefont {L.}~\bibnamefont {Khaykovich}}, \bibinfo
  {author} {\bibfnamefont {F.}~\bibnamefont {Chevy}},\ and\ \bibinfo {author}
  {\bibfnamefont {C.}~\bibnamefont {Salomon}},\ }\href
  {https://doi.org/10.1103/PhysRevA.87.063411} {\bibfield  {journal} {\bibinfo
  {journal} {Phys. Rev. A.}\ }\textbf {\bibinfo {volume} {87}},\ \bibinfo
  {pages} {063411} (\bibinfo {year} {2013})}\BibitemShut {NoStop}%
\bibitem [{\citenamefont {Lett}\ \emph {et~al.}(1989)\citenamefont {Lett},
  \citenamefont {Phillips}, \citenamefont {Rolston}, \citenamefont {Tanner},
  \citenamefont {Watts},\ and\ \citenamefont
  {Westbrook}}]{lettOpticalMolasses1989}%
  \BibitemOpen
  \bibfield  {author} {\bibinfo {author} {\bibfnamefont {P.~D.}\ \bibnamefont
  {Lett}}, \bibinfo {author} {\bibfnamefont {W.~D.}\ \bibnamefont {Phillips}},
  \bibinfo {author} {\bibfnamefont {S.~L.}\ \bibnamefont {Rolston}}, \bibinfo
  {author} {\bibfnamefont {C.~E.}\ \bibnamefont {Tanner}}, \bibinfo {author}
  {\bibfnamefont {R.~N.}\ \bibnamefont {Watts}},\ and\ \bibinfo {author}
  {\bibfnamefont {C.~I.}\ \bibnamefont {Westbrook}},\ }\href
  {https://doi.org/10.1364/josab.6.002084} {\bibfield  {journal} {\bibinfo
  {journal} {J. Opt. Soc. Am. B}\ }\textbf {\bibinfo {volume} {6}},\ \bibinfo
  {pages} {2084} (\bibinfo {year} {1989})}\BibitemShut {NoStop}%
\bibitem [{\citenamefont {Dedman}\ \emph {et~al.}(2001)\citenamefont {Dedman},
  \citenamefont {Baldwin},\ and\ \citenamefont
  {Colla}}]{dedmanFastSwitchingMagneticFieldsMOT2001}%
  \BibitemOpen
  \bibfield  {author} {\bibinfo {author} {\bibfnamefont {C.~J.}\ \bibnamefont
  {Dedman}}, \bibinfo {author} {\bibfnamefont {K.~G.~H.}\ \bibnamefont
  {Baldwin}},\ and\ \bibinfo {author} {\bibfnamefont {M.}~\bibnamefont
  {Colla}},\ }\href {https://doi.org/10.1063/1.1408935} {\bibfield  {journal}
  {\bibinfo  {journal} {Rev. Sci. Instrum.}\ }\textbf {\bibinfo {volume}
  {72}},\ \bibinfo {pages} {4055–4058} (\bibinfo {year} {2001})}\BibitemShut
  {NoStop}%
\bibitem [{\citenamefont {Matyjaśkiewicz}\ \emph {et~al.}(2008)\citenamefont
  {Matyjaśkiewicz}, \citenamefont {Szymańska},\ and\ \citenamefont
  {Góral}}]{matyjakiewiczProbingFermionicCondensates2008}%
  \BibitemOpen
  \bibfield  {author} {\bibinfo {author} {\bibfnamefont {S.}~\bibnamefont
  {Matyjaśkiewicz}}, \bibinfo {author} {\bibfnamefont {M.~H.}\ \bibnamefont
  {Szymańska}},\ and\ \bibinfo {author} {\bibfnamefont {K.}~\bibnamefont
  {Góral}},\ }\href {https://doi.org/10.1103/physrevlett.101.150410}
  {\bibfield  {journal} {\bibinfo  {journal} {Phys. Rev. Lett.}\ }\textbf
  {\bibinfo {volume} {101}},\ \bibinfo {pages} {150410} (\bibinfo {year}
  {2008})}\BibitemShut {NoStop}%
\bibitem [{\citenamefont {Kell}\ \emph {et~al.}(2021)\citenamefont {Kell},
  \citenamefont {Link}, \citenamefont {Breyer}, \citenamefont {Hoffmann},
  \citenamefont {K\"{o}hl},\ and\ \citenamefont
  {Gao}}]{kellCompactFastCoilFeshback2021}%
  \BibitemOpen
  \bibfield  {author} {\bibinfo {author} {\bibfnamefont {A.}~\bibnamefont
  {Kell}}, \bibinfo {author} {\bibfnamefont {M.}~\bibnamefont {Link}}, \bibinfo
  {author} {\bibfnamefont {M.}~\bibnamefont {Breyer}}, \bibinfo {author}
  {\bibfnamefont {A.}~\bibnamefont {Hoffmann}}, \bibinfo {author}
  {\bibfnamefont {M.}~\bibnamefont {K\"{o}hl}},\ and\ \bibinfo {author}
  {\bibfnamefont {K.}~\bibnamefont {Gao}},\ }\href
  {https://doi.org/10.1063/5.0049518} {\bibfield  {journal} {\bibinfo
  {journal} {Rev. Sci. Instrum.}\ }\textbf {\bibinfo {volume} {92}},\ \bibinfo
  {pages} {093202} (\bibinfo {year} {2021})}\BibitemShut {NoStop}%
\bibitem [{\citenamefont
  {Claussen}(2003)}]{ClaussenDynamicsBoseCondensate2003}%
  \BibitemOpen
  \bibfield  {author} {\bibinfo {author} {\bibfnamefont {N.~R.}\ \bibnamefont
  {Claussen}},\ }\emph {\bibinfo {title} {Dynamics of Bose-Einstein condensates
  near a Feshbach resonance in $^{85}$Rb}},\ \href@noop {} {\bibinfo {type}
  {Phd thesis}},\ \bibinfo  {school} {Faculty of the Graduate School of the
  University of Colorado}, \bibinfo {address} {Colorado} (\bibinfo {year}
  {2003}),\ \bibinfo {note} {available at
  \url{https://jila.colorado.edu/sites/default/files/2019-05/claussen_thesis.pdf}}\BibitemShut
  {NoStop}%
\bibitem [{\citenamefont {Stummer}(2007)}]{stummer2007}%
  \BibitemOpen
  \bibfield  {author} {\bibinfo {author} {\bibfnamefont {A.}~\bibnamefont
  {Stummer}},\ }\href
  {https://www.physics.utoronto.ca/~astummer/Archives/2007%20Mag-O-Matic%202/Mag-O-Matic%202.html}
  {\bibinfo {title} {{BEC} coil driver}} (\bibinfo {year} {2007}),\ \bibinfo
  {note} {accessed: June 2025}\BibitemShut {NoStop}%
\bibitem [{\citenamefont {Stummer}(2011)}]{stummer2011}%
  \BibitemOpen
  \bibfield  {author} {\bibinfo {author} {\bibfnamefont {A.}~\bibnamefont
  {Stummer}},\ }\href
  {https://www.physics.utoronto.ca/~astummer/Archives/2011%20Fesh-MOT%20Coils%20redux/Fesh-MOT%20Coils%20redux.html}
  {\bibinfo {title} {Fesh-{MOT} coils}} (\bibinfo {year} {2011}),\ \bibinfo
  {note} {accessed: June 2025}\BibitemShut {NoStop}%
\bibitem [{\citenamefont {Yang}\ \emph {et~al.}(2019)\citenamefont {Yang},
  \citenamefont {Xie}, \citenamefont {Ji}, \citenamefont {Wang}, \citenamefont
  {Zhang}, \citenamefont {Chen},\ and\ \citenamefont
  {Jiang}}]{YangBipolarCurrentDriver2019}%
  \BibitemOpen
  \bibfield  {author} {\bibinfo {author} {\bibfnamefont {Y.-M.}\ \bibnamefont
  {Yang}}, \bibinfo {author} {\bibfnamefont {H.-T.}\ \bibnamefont {Xie}},
  \bibinfo {author} {\bibfnamefont {W.-C.}\ \bibnamefont {Ji}}, \bibinfo
  {author} {\bibfnamefont {Y.-F.}\ \bibnamefont {Wang}}, \bibinfo {author}
  {\bibfnamefont {W.-Y.}\ \bibnamefont {Zhang}}, \bibinfo {author}
  {\bibfnamefont {S.}~\bibnamefont {Chen}},\ and\ \bibinfo {author}
  {\bibfnamefont {X.}~\bibnamefont {Jiang}},\ }\href
  {https://doi.org/10.1063/1.5046484} {\bibfield  {journal} {\bibinfo
  {journal} {Rev. Sci. Instrum.}\ }\textbf {\bibinfo {volume} {90}},\ \bibinfo
  {pages} {014701} (\bibinfo {year} {2019})}\BibitemShut {NoStop}%
\bibitem [{\citenamefont {Merkel}\ \emph {et~al.}(2019)\citenamefont {Merkel},
  \citenamefont {Thirumalai}, \citenamefont {Tarlton}, \citenamefont
  {Schäfer}, \citenamefont {Ballance}, \citenamefont {Harty},\ and\
  \citenamefont {Lucas}}]{MerkelMagFieldStabili2019}%
  \BibitemOpen
  \bibfield  {author} {\bibinfo {author} {\bibfnamefont {B.}~\bibnamefont
  {Merkel}}, \bibinfo {author} {\bibfnamefont {K.}~\bibnamefont {Thirumalai}},
  \bibinfo {author} {\bibfnamefont {J.~E.}\ \bibnamefont {Tarlton}}, \bibinfo
  {author} {\bibfnamefont {V.~M.}\ \bibnamefont {Schäfer}}, \bibinfo {author}
  {\bibfnamefont {C.~J.}\ \bibnamefont {Ballance}}, \bibinfo {author}
  {\bibfnamefont {T.~P.}\ \bibnamefont {Harty}},\ and\ \bibinfo {author}
  {\bibfnamefont {D.~M.}\ \bibnamefont {Lucas}},\ }\href
  {https://doi.org/10.1063/1.5080093} {\bibfield  {journal} {\bibinfo
  {journal} {Rev. Sci. Instrum.}\ }\textbf {\bibinfo {volume} {90}},\ \bibinfo
  {pages} {044702} (\bibinfo {year} {2019})}\BibitemShut {NoStop}%
\bibitem [{\citenamefont
  {Sabulsky}(2024)}]{sabulskyCurrentControllersOptimizing2024}%
  \BibitemOpen
  \bibfield  {author} {\bibinfo {author} {\bibfnamefont {D.~O.}\ \bibnamefont
  {Sabulsky}},\ }\href {https://doi.org/10.1063/5.0190625} {\bibfield
  {journal} {\bibinfo  {journal} {Rev. Sci. Instrum.}\ }\textbf {\bibinfo
  {volume} {95}},\ \bibinfo {pages} {081401} (\bibinfo {year}
  {2024})}\BibitemShut {NoStop}%
\bibitem [{\citenamefont {Eigen}(2019)}]{eigenExploringBoseGasesAuxCoils2019}%
  \BibitemOpen
  \bibfield  {author} {\bibinfo {author} {\bibfnamefont {C.}~\bibnamefont
  {Eigen}},\ }\emph {\bibinfo {title} {Exploring Interacting Bose Gases in and
  out of Equilibrium}},\ \href@noop {} {\bibinfo {type} {Phd thesis}},\
  \bibinfo  {school} {The Cavendish laboratory, University of Cambridge},
  \bibinfo {address} {Cambridge} (\bibinfo {year} {2019}),\ \bibinfo {note}
  {available at
  \url{https://www.repository.cam.ac.uk/handle/1810/299008}}\BibitemShut
  {NoStop}%
\bibitem [{\citenamefont {Ahmed}\ \emph {et~al.}(2019)\citenamefont {Ahmed},
  \citenamefont {Altiere}, \citenamefont {Andalib}, \citenamefont {Barnes},
  \citenamefont {Bell}, \citenamefont {Bidinosti}, \citenamefont {Bylinsky},
  \citenamefont {Chak}, \citenamefont {Das}, \citenamefont {Davis},
  \citenamefont {Fischer}, \citenamefont {Franke}, \citenamefont {Gericke},
  \citenamefont {Giampa}, \citenamefont {Hahn}, \citenamefont {Hansen-Romu},
  \citenamefont {Hatanaka}, \citenamefont {Hayamizu}, \citenamefont {Jamieson},
  \citenamefont {Jones}, \citenamefont {Katsika}, \citenamefont {Kawasaki},
  \citenamefont {Kikawa}, \citenamefont {Klassen}, \citenamefont {Konaka},
  \citenamefont {Korkmaz}, \citenamefont {Kuchler}, \citenamefont
  {Kurchaninov}, \citenamefont {Lang}, \citenamefont {Lee}, \citenamefont
  {Lindner}, \citenamefont {Madison}, \citenamefont {Mammei}, \citenamefont
  {Mammei}, \citenamefont {Martin}, \citenamefont {Matsumiya}, \citenamefont
  {Miller}, \citenamefont {Momose}, \citenamefont {Picker}, \citenamefont
  {Pierre}, \citenamefont {Ramsay}, \citenamefont {Rao}, \citenamefont
  {Rawnsley}, \citenamefont {Rebenitsch}, \citenamefont {Schreyer},
  \citenamefont {Sidhu}, \citenamefont {Vanbergen}, \citenamefont {van Oers},
  \citenamefont {Watanabe},\ and\ \citenamefont
  {Yosifov}}]{AhmedFastMagnet2019}%
  \BibitemOpen
  \bibfield  {author} {\bibinfo {author} {\bibfnamefont {S.}~\bibnamefont
  {Ahmed}}, \bibinfo {author} {\bibfnamefont {E.}~\bibnamefont {Altiere}},
  \bibinfo {author} {\bibfnamefont {T.}~\bibnamefont {Andalib}}, \bibinfo
  {author} {\bibfnamefont {M.~J.}\ \bibnamefont {Barnes}}, \bibinfo {author}
  {\bibfnamefont {B.}~\bibnamefont {Bell}}, \bibinfo {author} {\bibfnamefont
  {C.~P.}\ \bibnamefont {Bidinosti}}, \bibinfo {author} {\bibfnamefont
  {Y.}~\bibnamefont {Bylinsky}}, \bibinfo {author} {\bibfnamefont
  {J.}~\bibnamefont {Chak}}, \bibinfo {author} {\bibfnamefont {M.}~\bibnamefont
  {Das}}, \bibinfo {author} {\bibfnamefont {C.~A.}\ \bibnamefont {Davis}},
  \bibinfo {author} {\bibfnamefont {F.}~\bibnamefont {Fischer}}, \bibinfo
  {author} {\bibfnamefont {B.}~\bibnamefont {Franke}}, \bibinfo {author}
  {\bibfnamefont {M.~T.~W.}\ \bibnamefont {Gericke}}, \bibinfo {author}
  {\bibfnamefont {P.}~\bibnamefont {Giampa}}, \bibinfo {author} {\bibfnamefont
  {M.}~\bibnamefont {Hahn}}, \bibinfo {author} {\bibfnamefont {S.}~\bibnamefont
  {Hansen-Romu}}, \bibinfo {author} {\bibfnamefont {K.}~\bibnamefont
  {Hatanaka}}, \bibinfo {author} {\bibfnamefont {T.}~\bibnamefont {Hayamizu}},
  \bibinfo {author} {\bibfnamefont {B.}~\bibnamefont {Jamieson}}, \bibinfo
  {author} {\bibfnamefont {D.}~\bibnamefont {Jones}}, \bibinfo {author}
  {\bibfnamefont {K.}~\bibnamefont {Katsika}}, \bibinfo {author} {\bibfnamefont
  {S.}~\bibnamefont {Kawasaki}}, \bibinfo {author} {\bibfnamefont
  {T.}~\bibnamefont {Kikawa}}, \bibinfo {author} {\bibfnamefont
  {W.}~\bibnamefont {Klassen}}, \bibinfo {author} {\bibfnamefont
  {A.}~\bibnamefont {Konaka}}, \bibinfo {author} {\bibfnamefont
  {E.}~\bibnamefont {Korkmaz}}, \bibinfo {author} {\bibfnamefont
  {F.}~\bibnamefont {Kuchler}}, \bibinfo {author} {\bibfnamefont
  {L.}~\bibnamefont {Kurchaninov}}, \bibinfo {author} {\bibfnamefont
  {M.}~\bibnamefont {Lang}}, \bibinfo {author} {\bibfnamefont {L.}~\bibnamefont
  {Lee}}, \bibinfo {author} {\bibfnamefont {T.}~\bibnamefont {Lindner}},
  \bibinfo {author} {\bibfnamefont {K.~W.}\ \bibnamefont {Madison}}, \bibinfo
  {author} {\bibfnamefont {J.}~\bibnamefont {Mammei}}, \bibinfo {author}
  {\bibfnamefont {R.}~\bibnamefont {Mammei}}, \bibinfo {author} {\bibfnamefont
  {J.~W.}\ \bibnamefont {Martin}}, \bibinfo {author} {\bibfnamefont
  {R.}~\bibnamefont {Matsumiya}}, \bibinfo {author} {\bibfnamefont
  {E.}~\bibnamefont {Miller}}, \bibinfo {author} {\bibfnamefont
  {T.}~\bibnamefont {Momose}}, \bibinfo {author} {\bibfnamefont
  {R.}~\bibnamefont {Picker}}, \bibinfo {author} {\bibfnamefont
  {E.}~\bibnamefont {Pierre}}, \bibinfo {author} {\bibfnamefont {W.~D.}\
  \bibnamefont {Ramsay}}, \bibinfo {author} {\bibfnamefont {Y.-N.}\
  \bibnamefont {Rao}}, \bibinfo {author} {\bibfnamefont {W.~R.}\ \bibnamefont
  {Rawnsley}}, \bibinfo {author} {\bibfnamefont {L.}~\bibnamefont
  {Rebenitsch}}, \bibinfo {author} {\bibfnamefont {W.}~\bibnamefont
  {Schreyer}}, \bibinfo {author} {\bibfnamefont {S.}~\bibnamefont {Sidhu}},
  \bibinfo {author} {\bibfnamefont {S.}~\bibnamefont {Vanbergen}}, \bibinfo
  {author} {\bibfnamefont {W.~T.~H.}\ \bibnamefont {van Oers}}, \bibinfo
  {author} {\bibfnamefont {Y.~X.}\ \bibnamefont {Watanabe}},\ and\ \bibinfo
  {author} {\bibfnamefont {D.}~\bibnamefont {Yosifov}},\ }\href
  {https://doi.org/10.1103/PhysRevAccelBeams.22.102401} {\bibfield  {journal}
  {\bibinfo  {journal} {Phys. Rev. Accel. Beams}\ }\textbf {\bibinfo {volume}
  {22}},\ \bibinfo {pages} {102401} (\bibinfo {year} {2019})}\BibitemShut
  {NoStop}%
\bibitem [{\citenamefont
  {Bechhoefer}(2005)}]{bechhoeferFeedbackPhysicistsTutorial2005}%
  \BibitemOpen
  \bibfield  {author} {\bibinfo {author} {\bibfnamefont {J.}~\bibnamefont
  {Bechhoefer}},\ }\href {https://doi.org/10.1103/RevModPhys.77.783} {\bibfield
   {journal} {\bibinfo  {journal} {Rev. Mod. Phys.}\ }\textbf {\bibinfo
  {volume} {77}},\ \bibinfo {pages} {783} (\bibinfo {year} {2005})}\BibitemShut
  {NoStop}%
\bibitem [{\citenamefont {Ogata}(2010)}]{ogataModernControlEngineering2010}%
  \BibitemOpen
  \bibfield  {author} {\bibinfo {author} {\bibfnamefont {K.}~\bibnamefont
  {Ogata}},\ }\href
  {https://www.pearson.com/en-us/subject-catalog/p/modern-control-engineering/P200000003521/9780137551064}
  {\emph {\bibinfo {title} {Modern Control Engineering}}},\ \bibinfo {edition}
  {5th}\ ed.\ (\bibinfo  {publisher} {Pearson},\ \bibinfo {address} {Boston},\
  \bibinfo {year} {2010})\BibitemShut {NoStop}%
\bibitem [{\citenamefont {Horowitz}\ and\ \citenamefont
  {Hill}(2024)}]{horowitzArtElectronics2024}%
  \BibitemOpen
  \bibfield  {author} {\bibinfo {author} {\bibfnamefont {P.}~\bibnamefont
  {Horowitz}}\ and\ \bibinfo {author} {\bibfnamefont {W.}~\bibnamefont
  {Hill}},\ }\href {https://artofelectronics.net/} {\emph {\bibinfo {title}
  {The Art of Electronics}}},\ \bibinfo {edition} {third edition, 20th printing
  with corrections}\ ed.\ (\bibinfo  {publisher} {Cambridge University Press},\
  \bibinfo {address} {Cambridge, New York},\ \bibinfo {year}
  {2024})\BibitemShut {NoStop}%
\bibitem [{IPC(1998)}]{IPC2221}%
  \BibitemOpen
  \href@noop {} {\emph {\bibinfo {title} {Generic Standard on Printed Board
  Design (IPC-2221)}}},\ \bibinfo {organization} {Institute for Printed
  Circuits (IPC)},\ \bibinfo {address} {Bannockburn, IL, USA} (\bibinfo {year}
  {1998})\BibitemShut {NoStop}%
\bibitem [{\citenamefont {{Kepco, Inc.}}(2023)}]{kepcoBOP}%
  \BibitemOpen
  \bibfield  {author} {\bibinfo {author} {\bibnamefont {{Kepco, Inc.}}},\
  }\href@noop {} {\bibinfo {title} {Bop models optimized for inductive
  loads}},\ \bibinfo {howpublished}
  {\url{https://www.kepcopower.com/bop-ind.htm}} (\bibinfo {year} {2023}),\
  \bibinfo {note} {accessed: June 2025}\BibitemShut {NoStop}%
\bibitem [{rep(2025)}]{repo}%
  \BibitemOpen
  \href@noop {} {\bibinfo {title} {{Fast Coil Driver Workfiles Repository}}},\
  \bibinfo {howpublished} {\url{https://github.com/ifquetzal/coildriver}}
  (\bibinfo {year} {2025}),\ \bibinfo {note} {[Online; accessed
  1-March-2025]}\BibitemShut {NoStop}%
\end{thebibliography}%

\end{document}